# A Synergized Pulsing-Imaging Network (SPIN)

Qing Lyu[1], Tao Xu[2], Hongming Shan[1], Ge Wang[1]*

1 Biomedical Imaging Center, Department of Biomedical Engineering, Rensselaer Polytechnic Institute, Troy, New York, USA, 12180

2 School of Software and Microelectronics, Northwestern Polytechnical University, Xi'an, Shaanxi, China, 710072.

lyuq@rpi.edu; xutao@nwpu.edu.cn; shanh@rpi.edu; ge-wang@ieee.org

*Abstract* — **Currently, the deep neural network is the mainstream for machine learning, and being actively developed for biomedical imaging applications with an increasing emphasis on tomographic reconstruction for MRI, CT, and other imaging modalities. Multiple deep-learning-based approaches were applied to MRI image reconstruction from *k*-space samples to final images. Each of these studies assumes a given pulse sequence that produces incomplete and/or inconsistent data in the Fourier space, and targets a trained neural network that recovers an underlying image as close as possible to the ground truth. For the first time, in this paper we view data acquisition and the image reconstruction as the two key parts of an integrated MRI process, and optimize both the pulse sequence and the reconstruction scheme seamlessly in the machine learning framework. Our pilot simulation results show an exemplary embodiment of our new MRI strategy. Clearly, this work can be extended to other imaging modalities and their combinations as well, such as ultrasound imaging, and also potentially "simultaneous emission-transmission tomography" aided by polarized radiotracers.**

## I. INTRODUCTION

Computer vision and image analysis are both great examples of successes with machine learning especially deep learning. While computer vision and image analysis primarily handle existing images and extract their features, tomographic reconstruction produces internal images from indirect data or features of underlying images. An overall perspective on deep imaging was published in [1], which underlined that the combination of medical imaging and deep learning promises to empower not only image analysis but also image reconstruction, and suggested to develop image reconstruction techniques using machine learning techniques. The hope is to enable intelligent utilization of domain knowledge in terms of big data, innovative approaches for image reconstruction, and superior performance in clinical and preclinical applications. Recently, deep learning techniques are being actively explored for tomographic reconstruction by multiple groups worldwide, with encouraging results and strong momentums. We believe that deep reconstruction is a next major target of deep learning.

As mentioned in the above perspective [1], "*In many imaging modalities, data quality and image metrics are complicated by multiple factors, such as imaging geometry, patient placement, sensor calibration, and so on. Prior information about these characteristics was taken into account, to different degrees, in existing reconstruction approaches but it is not straightforward within the deep learning framework.*" "*To the first order approximation, a majority of medical imaging algorithms have Fourier or wavelet transform related versions, and could be helped by some common deep networks. For nonlinear imaging models, deep imaging should be a better strategy, given the nonlinear nature of deep networks. While the multimodality imaging trend promotes a system-level integration, deep imaging might be a unified information theoretic framework or a meta-solution to support either individual or hybrid scanners.*"

Taking MRI as an example, the common imaging model is a Fourier formulation. With a homogeneous background magnetic field, all the magnetization vectors associated with individual pixels/voxels are aligned up along the direction of the background field. Then, a pulse sequence is applied to perturb these vectors and produce non-zero components on the place perpendicular to the main field. These in-plane vectors generate alternating electromagnetic fields in nearby coils to produce so-called free induction decay (FID) signals. The recorded data are approximated as samples, which are known as *k*-space data, of the Fourier transform of the patient or animal to be tomographically reconstructed. If the Fourier space is fully sampled by the pulse sequence, an MRI image can be directly reconstructed using the inverse Fourier transform.

Up to now, deep learning has only been applied to *k*-space MRI data. Several neural networks were recently proposed for under-sampled MRI. Typically, a deep neural network contains many parameters trainable with a large set of fully-sampled high-quality *k*-space datasets. For example, a deep learning method for faster MRI was reported in [2]. First, *k*-space data were selected via sub-Nyquist sampling in the phase-encoding direction. To unravel image folding, some low-frequency *k*-space data were included. Numerous results suggested a promising performance of the proposed method with 29% of fully-sampled *k*-space data. As another example, a transfer-learning approach was proposed to address training data scarcity for accelerated MRI [3]. The neural network was trained using samples from natural images on *ImageNet*, and fine-tuned with only tens of MR images of interest (*T1*- or *T2*-weighted images). The results demonstrated that the ImageNet-trained network was nearly identical to that trained with thousands of MRI images, outperforming compressed sensing reconstruction quality.

However, the above-described Fourier-space-based approaches are far from perfect in many important scenarios. Even if the MRI scanner is ideal in its specifications, the patient or animal to be reconstructed is constrained by the complicating physical and physiological factors. Generally speaking, the MRI signal is not strong, since only a tiny imbalance of up and down spins contributed to the net FID signal. Also, during data acquisition, the signal undergoes *T2* defocusing, which is ignored in the current Fourier formulation. Furthermore, the moving spins are either assumed to be stationary or only modeled in terms of low-order moments. All these and other approximations could lead to rapidly increased errors/biases when the overall data acquisition time is greatly shortened for fast MRI tasks such as in challenging cardiovascular or brain studies. Our main hypothesis is that for the best MRI performance, the conventional *k*-space formulation should be upgraded so that the data acquisition and image reconstructed steps are integrated in the machine learning framework. In other words, we should learn not only a network for image reconstruction but also a pulse sequence that facilitates task-specific image reconstruction and subsequent analysis, and should integrate these two parts into a whole process for systematic optimization.

In this paper, we propose to synergize the pulse sequence design and the associated imaging method into a deep neural network, abbreviated as "SPIN" for the Synergized Pulsing-Imaging Network. With our SPIN, in principle any hand-crafted pulse sequence can be re-discovered including spin-echo (SE), gradient-echo (GE), echo planar imaging (EPI) and even MR fingerprint (MRF) sequences [4], and non-existing pulse sequences could be invented by machine learning. Through our SPIN, not only the information content of raw MRI data can be maximized in a task-specific fashion but also resultant MRI images can be optimally reconstructed in a data-driven manner.

The rest of the paper is organized as follows. In the next section, the SPIN approach is illustrated in an exemplary setting. An in-house numerical simulator is described to manipulate the magnetization vectors inside a patient or animal and solve the Bloch equation to synthesize data, which may or may not be casted in the Fourier space accurately. Then, an overall workflow is presented to optimize the MRI image quality with respect to both pulse sequence parameters and neural network parameters. In the third section, typical simulation results are given to illustrate key steps of our workflow in a toy world consisting of various mathematical brain phantoms equipped with typical proton density ($\rho$), *T1* and *T2* values. In the last section, relevant issues and further topics are discussed.

## II. Methodology

Our deep learning approach "SPIN" targets a data-driven mapping from an object to sampled MRI data and then to an MRI image of the object, without necessarily interpreting the data as Fourier coefficients and performing the inverse Fourier transform. The overall idea is illustrated in Figure 1. The two key components are (1) data acquisition with a pulse sequence and (2) image reconstruction with a neural network. First, from real objects or synthesized true images, raw MRI datasets can be generated according to a pulse sequence, which can be gradually refined in the training process after initialization. The data can be fed into a neural network with a sufficient expressing capability. The reconstruction neural network needs to be trained as well. The loss function is driven by the difference between the truth and the output of the reconstruction neural network, and used to guide the optimization of both the pulse sequence and the reconstruction network. Such an optimization procedure can be conducted in various ways. In this pilot study, we first trained the neural network with data generated from a standard spin echo pulse sequence, and then optimized the spin echo pulse sequence through the loss function. This alternating optimization process is to demonstrate how the SPIN approach deals with both data acquisition and image reconstruction in one embodiment, and many other ways are certainly possible.

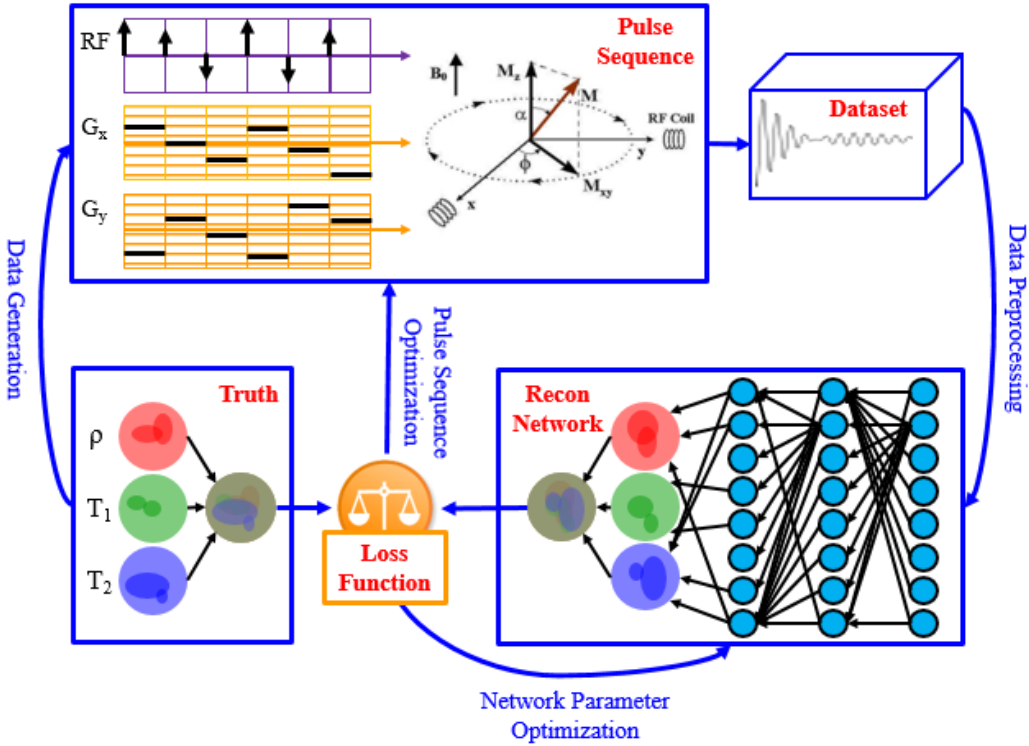

*Figure 1. Concept of the proposed SPIN strategy*. This design can be significantly refined with respect to imaging system setups, computational steps, and specific applications.

### A. Data Generation

2D Shepp-Logan-type phantoms were created for training and testing the neural network. These phantoms represent brain tissues like scalp, bone, CSF, gray and white matter and tumors. The parameters of the phantoms were set in reference to [5]. Some ellipses' half-axis parameters were enlarged to show the corresponding components more clearly. The Shepp-Logan phantom specifications used in this paper for SPIN-based MRI is shown in Table 1. Each phantom can be viewed an idealized brain organoid with $\rho$, $T1$ and $T2$ components in registered three 64×64 matrices. When designing the phantoms, positions of the 8 tissue components (one gray matter, two CSF regions, and five tumors) were randomly positioned in the central region, and there was no overlap between any two of these components. Totally, 50,000 phantoms were made as the training dataset, and another 1,000 phantoms were created for testing, focusing on the flip angle optimization.

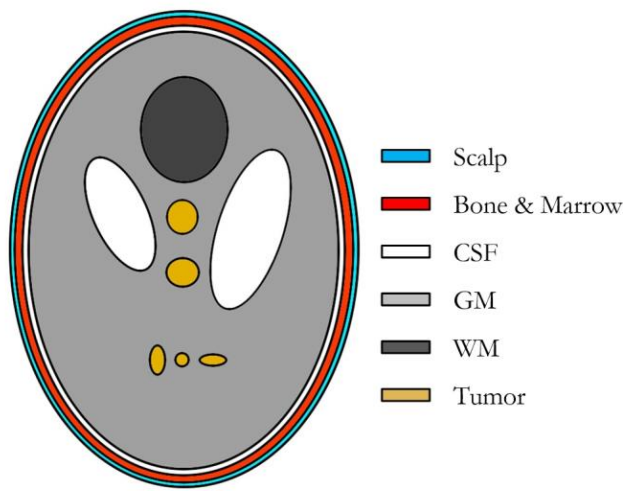

*Figure 2. Illustration of the Shepp-Logan-type phantom.*

*Table 1. Shepp-Logan Phantom Specifications for SPIN-based MRI.*

| Ellipsoid | Center | | Half-Axis | | Angle (degree) | ρ | T1 (ms) | T2 (ms) | Tissue |
|---|---|---|---|---|---|---|---|---|---|
| | x | y | a | b | | | | | |
| 1 | 0 | 0 | 0.72 | 0.95 | 0 | 0.8 | 324 | 70 | Scalp |
| 2 | 0 | 0 | 0.69 | 0.92 | 0 | 0.12 | 533 | 50 | Bone & Marrow |
| 3 | 0 | -0.0184 | 0.66 | 0.87 | 0 | 0.98 | 4200 | 1990 | CSF |
| 4 | 0 | -0.0184 | 0.63 | 0.84 | 0 | 0.745 | 857 | 100 | GM |
| 5 | - | - | 0.35 | 0.13 | -72 | 0.98 | 4200 | 1990 | CSF |
| 6 | - | - | 0.28 | 0.09 | 72 | 0.98 | 4200 | 1990 | CSF |
| 7 | - | - | 0.18 | 0.22 | 0 | 0.617 | 583 | 80 | WM |
| 8 | - | - | 0.08 | 0.08 | 0 | 0.95 | 926 | 100 | Tumor |
| 9 | - | - | 0.08 | 0.05 | 0 | 0.95 | 926 | 100 | Tumor |
| 10 | - | - | 0.08 | 0.05 | -90 | 0.95 | 926 | 100 | Tumor |
| 11 | - | - | 0.08 | 0.08 | 0 | 0.95 | 926 | 100 | Tumor |
| 12 | - | - | 0.08 | 0.08 | 0 | 0.95 | 926 | 100 | Tumor |

A classical spin echo pulse sequence is shown in Figure 2. The repetition time (TR) and the echo time (TE) are 10s and 40ms respectively. There was a radio frequency (RF) pulse in the beginning (t=0ms) to flip the magnetization vector by a certain angle. Typically, this flip angle is between 0° and 180° determined by the specific RF pulse. For a standard spin echo pulse sequence, this flip angle is 90°. The second RF pulse was produced at half of the echo time (t = 20ms), and the resultant flip angle is typically 180°. There was a gradient magnetic field $G_x$ during 16.8~20ms, 36.8~43.2ms and $G_y$ between 16.8~20ms. The Bloch equation governs the MRI data generation. Without loss of generality, we restricted our data generation process to an idealized 2D case over the spin-echo period of a limited length. The spin-echo period was discretized into a number of identical intervals or time units. At the beginning (t=0s), we assumed that the magnetization vector $M_0$ for every pixel was well aligned with the direction of the background magnetic field. During the data acquisition, the pixel-specific magnetic vector as a function of time can be calculated in terms of $M_x$, $M_y$, and $M_z$ step by step in a fine time resolution, say 0.1ms, according to the Bloch equation.

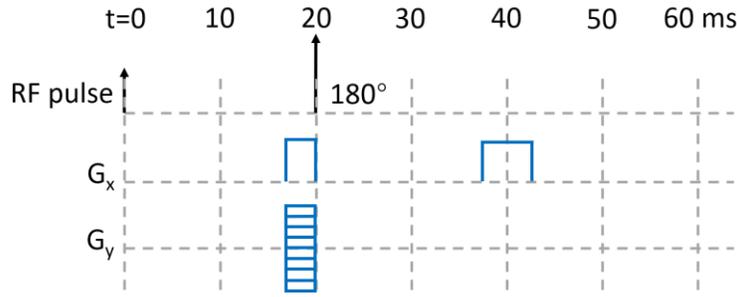

*Figure 3. Discretized spin-echo pulse sequence that be randomly initialized and iteratively refined.*

In the rotating reference frame, after a rapid flip of the magnetization vector, $M_{xy}(t)$ and $M_z(t)$ over the first time interval can be expressed as follows:

$$M_{xy}(t) = M_{xy}(0)e^{\frac{-t}{T_2}} = M_x(0)e^{\frac{-t}{T_2}} + iM_y(0)e^{\frac{-t}{T_2}} \tag{1}$$

$$M_z(t) = M_z(0)e^{\frac{-t}{T_1}} + M_0(1 - e^{\frac{-t}{T_1}}) \tag{2}$$

where $T_2$ should be $T_2^*$ if the pixel size is not sufficiently small. With the $G_x$ and $G_y$ effects, the pixel-specific $M_{xy}(t)$ vectors will rotate around the z-axis at pixel-specific frequencies $w = \gamma(G_x x + G_y y)$. That is,

$$M_{xy}(t) = M_{xy}(0)e^{\frac{-t}{T_2}}e^{-i\gamma(G_x x + G_y y)} \tag{3}$$

and the resultant complex-valued FID signal is recorded as $s_x(t)$ and $s_y(t)$ as the real and imaginary parts respectively. Ignoring the scaling factor, we have

$$s_x(t) = Real(\sum M_{xy}(t) \cdot \rho) \tag{4}$$

$$s_y(t) = Imag(\sum M_{xy}(t) \cdot \rho) \tag{5}$$

Then, we can sample $s_x(t)$ and $s_y(t)$ (say, every 0.1ms) to generate data over the first time unit.

By the end of the first time interval, the pixel-specific magnetization vectors will generally not align with the z axis. Hence, at the beginning of the 2nd time interval, the magnetization vector has a new initial state ($M'_x(0)$, $M'_y(0)$, $M'_z(0)$) different from that for the 1st time interval. Specifically,

$$M'_x(0) = M_x(0)e^{\frac{-\Delta}{T_2}}\cos(\gamma(xG_x + yG_y)\Delta) \tag{6}$$

$$M'_y(0) = M_y(0)e^{\frac{-\Delta}{T_2}}\sin(\gamma(xG_x + yG_y)\Delta) \tag{7}$$

$$M'_z(0) = M_z(0)e^{\frac{-\Delta}{T_1}} + M_0(1 - e^{\frac{-\Delta}{T_1}}) \tag{8}$$

where $\Delta$ is the time interval length. With the initial condition specified by Eqs. (6-8), The Bloch equations can be solved again over the 2nd time interval. Then, this procedure can be repeated for each of the rest time intervals to generate raw MRI data, as shown in Figure 3.

When down-sampling the signal according to the *k*-space theorem (which is an approximation, as mentioned earlier), we have

$$\Delta k_x = \frac{\gamma}{2\pi}G_x \Delta t = \frac{1}{2x_{max}} \tag{8}$$

$$x_{max} = \frac{FOV_x}{2} \tag{9}$$

$$\Delta k_y = \frac{\gamma}{2\pi}G_y T_{ph} = \frac{1}{2y_{max}} \tag{10}$$

$$G_y = n g_y \tag{11}$$

$$y_{max} = \frac{FOV_y}{2} \tag{12}$$

where $FOV_x = FOV_y = 64$mm, $\gamma = 42.576$MHz, $\Delta t = 0.1$ms, $G_x = 0.23\ T/m$, $T_{ph} = 6.4\ ms$, $g_y = 0.0072T/m$, and *n* is the index for phase-encoding.

For a given phantom, data were first synthesized over 64 standard spin-echo periods. In each period, 64 data points were sampled at the frequency of 10,000 Hz. Data sampled from each phantom were recorded in a 64 × 64 complex-valued matrix. The process of data generation was implemented in Python 3.6 on the CUDA platform. Generating data over 64 phase-encoding periods took about 2 seconds. Then, the data were compromised with additive Gaussian noise of mean 0 and standard deviation 2.5% of the max signal magnitude. Figure 4 shows the reconstructed MRI images using the inverse Fourier transform directly.

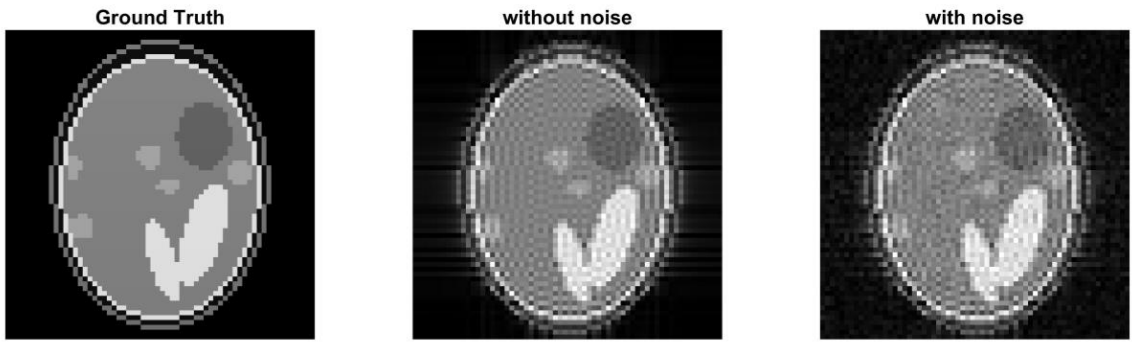

*Figure 4. Reconstructed images using the inverse Fourier Transform directly. Left: The ground truth; Middle: the image without noise; and Right: the image with noise.*

### B. *Image Reconstruction*

After the direct Fourier inversion as the benchmark, we utilized a convolutional neural network (CNN) [6]. The architecture of the CNN is shown in Figure 5. It consisted of two fully-connected layers, two convolutional layers, and a deconvolutional layer. The input was the raw MRI data of standard spin-echo pulse sequence (90º and 180º flip angles) sampled at 10,000 Hz (see II.A). Before the data were fed to the neural network, they were first vectorized from the original 64 × 64 complex-valued matrix to a 64 × 64 × 2 real number vector (splitting each complex number into two real numbers representing real and imaginary parts respectively). The ground truth for each phantom was a 64 × 64 matrix from the phantom T1, T2 and proton density values by calculating

$$s = M_0(1 - e^{-TR/T_1})e^{-t/T_2} \cdot \rho \qquad (13)$$

where *t* is the sampling time (between 36.8ms and 43.2ms over each spin-echo period).

The loss function was a combination of the mean-squared-error and the $L_1$-norm penalty, with a relaxation factor λ = 0.0001. This $L_1$-norm penalty was applied to the feature map of the final hidden layer. The RMSProp algorithm [7] was used with mini-batches of size 100, momentum 0.0, and decay 0.9. The learning rate was 0.0005 for the first 250 epochs, and then divided by 1.01 every epoch. The neural network was implemented in TensorFlow [8] on NVIDIA GTX 1080Ti. The neural network was initialized using the method described in [9]. The training process took 300 epochs and about 5 hours.

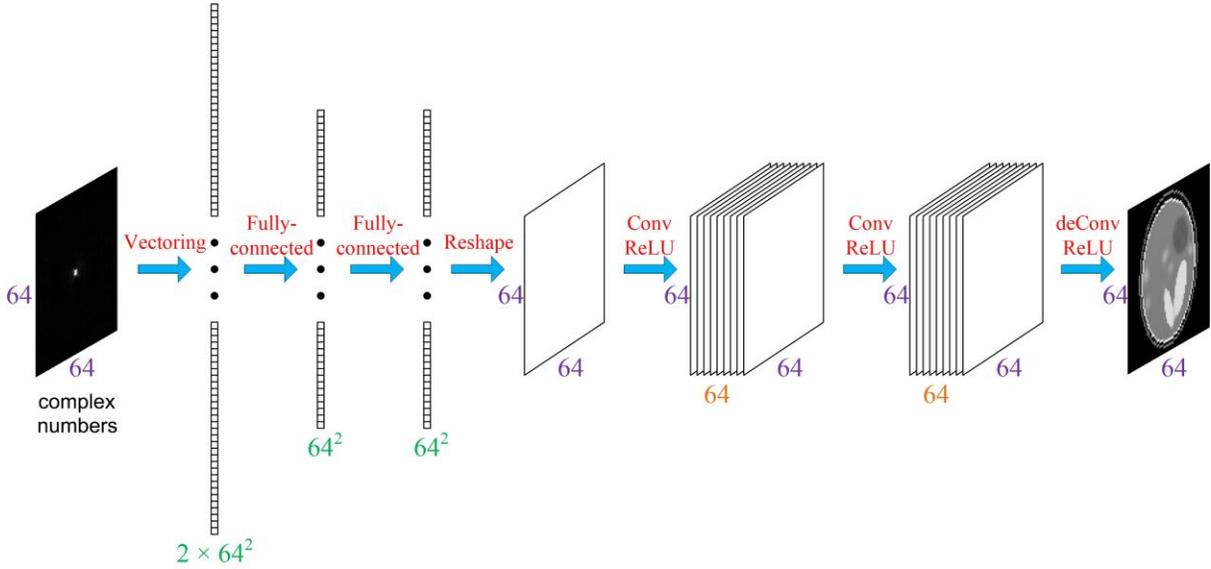

*Figure 5. Neural network used for MRI image reconstruction.*

### C. Parameter Optimization

In the SPIN scheme, there are two sets of parameters to be trained. The first set of parameters is for the pulse sequence, involving angle-flip RF pulses and gradient signals $G_x$ and $G_y$ for each time interval. These parameters specify how exactly raw MRI data are generated. The second set of parameters defines the neural network so that it can reconstruct an MRI image up to a high quality.

For SPIN-based optimization, at the beginning we could randomly initialize all the parameters. Then, MRI raw data can be obtained according to the Bloch equation. After that, the data are fed into the neural network. Through a number of epochs, the neural network is trained to have the second set of parameters updated so that the initial value of the loss function $L_0$ is obtained. Then, the first set of parameters can be updated. This two-step process is a form of alternating iterative optimization, and was chosen here to demonstrate the feasibility of SPIN. However, other SPIN implementations are certainly possible.

Specifically, each element in the first set of parameters is perturbed by a small quantity δ to estimate the partial derivative of the loss function with respect to that parameter. For example, the RF pulse flip angle $\alpha_1$ for the 1st time interval can be changed to $\alpha_1' = \alpha_1 + \delta$, and then the CNN can be re-trained to obtain a new value of the loss function $L_1$. Then, the corresponding partial derivative can be estimated as

$$\frac{\partial F}{\partial \alpha_1'} = \frac{L_1 - L_0}{\delta} \tag{13}$$

In the same way, all the other partial derivatives can be obtained to form the gradient vector $(\frac{\partial F}{\partial \alpha_1'}, \frac{\partial F}{\partial G_{x1}'}, \frac{\partial F}{\partial G_{y1}'}, ...)$, which can be used to improve the current pulse sequence. Along the steepest descent direction at an appropriate learning rate, for example, $\eta = 0.001$, the current pulse sequence can be updated to $(\alpha_1 + \eta \frac{\partial F}{\partial \alpha_1'}, G_{x1} + \frac{\partial F}{\partial G_{x1}'}, ...)$. By now, the 1st learning cycle is finished. Such a learning cycle must be repeated multiple times with respect to the first and second sets of parameters, and with respect to each of the ground truth objects/images. By doing so, the loss function will be gradually minimized, hopefully in most of cases, until a stopping criterion is satisfied.

For an initial demonstration, here we just implemented the optimization of the first flip angle in the spin-echo pulse sequence. All other parameters except the first flip angle in the first set of parameters were fixed for simplicity. Also, we optimized the second set of parameters by training the reconstruction neural network with MRI raw data generated from standard spin-echo pulse sequence as our initial reconstruction network. Then, we set the first flip angle to 30º and obtained new raw MRI data. After that, MRI images were reconstructed through the trained neural networks, with the loss function defined as the mean squared error ($L_0$) of the reconstructed MRI image and the corresponding ground truth. Furthermore, we changed the first flip angle to 30º plus δ, and similarly obtained a new mean squared error ($L_1$). According to the two mean squared errors, a new flip angle can be found using the aforementioned gradient descent method. After a sufficiently many learning cycles, the flip angle was optimized to produce satisfactory MRI image reconstructions.

## III. NUMERICAL RESULTS

A world of toy phantoms was created for this pilot study, with a representative phantom shown in Figure 2. Each phantom was discretized into $64 \times 64$ pixels, each of which has $ρ$, $T1$ and $T2$ components. Collectively, these three components define biologically meaningful brain features: CSF, gray matter, white matter, and tumors. We assumed that the width and height of phantom were both 64mm. Each spin-echo pulse sequence lasted 10s, with 100,000 time units (Each time unit was set to 0.1ms). In principle, there could be an RF pulse at the beginning of each time unit and the resonance frequency determined by $B_0$ with a strength $B_1$. The flip angle could be randomly set between 0-180˚. However, in this demonstration we only focused on the optimization for the 90-degree flip angle.

Using the inverse Fourier Transform directly, the reconstructed image had significant Gibbs artifacts as shown in Figure 4. This was due to the limited number of data points in the $k$ space. Adding noise aggravated the artifacts and made it harder to recognize tissue components in the phantom. In Figure 6, when the flip angle was far from 90º (the original and the first a few iterations), the reconstructed MRI images were dim, and their average pixel intensities were low, due to the weak MRI FID signals. However, when the flip angle was gradually improved towards 90º, the MRI signals became stronger, and the reconstructed images were in better quality, leading to the clearer definitions of the brain tissue components.

Without using the trained convolutional neural network to reconstruct MRI image, the MRI images through the Fourier inversion had significant Gibbs artifacts and noises, as shown in Figures 4 and 6. In contrast, using the proposed SPIN method that uses the trained convolutional neural network to reconstruct MRI images, Gibbs artifacts and noises were greatly suppressed in the reconstructed images as shown in Figure 7. This showcased the power of the neural network in noise reduction and artifact correction, which is consistent to other reports such as [10].

In the first several iterations, the reconstructed MRI images contained some "fake spots" suggesting the brain features that are not in existence. These spots might be produced from the added noises. When the first flip angle was small, the signal strength was weak. The added noises could influence the reconstructed image more, forming image artifacts. As the optimization process went more deeply, those fake spots were eliminated, and the true brain tissue features were kept in the reconstructed images. Quantitatively, with the increasing number of optimization iterations, the structural similarity index (SSIM) and the peak signal-to-noise ratio (PSNR) became gradually improved, as shown in Figure 8. Before the flip angle optimization, the average SSIM of the 1,000 reconstructed MRI image and the average PSNR were 0.932 and 28.3 respectively. After the first 7 iterations, there measures were refined to 0.945 and 29.2 respectively. Although the trained CNN reconstruction is powerful in alleviating artifacts and noises, the brain tissues in the reconstructed MRI images look still a bit blurry. This is mainly due to the limited number of the phantom pixels, and can be effectively addressed by increasing the size of the phantom.

Figure 9 shows the variation of the mean squared error between the reconstructed images and their corresponding phantom ground truth, which implies that through iterations the mean squared error was greatly decreased. The flip angle was optimized to become close to 90º that is the optimal flip angle for the spin-echo pulse. Quantitatively, the value of the mean squared error was reduced from 0.001496 associated with the randomly-selected flip angle 30º before the optimization to 0.001362 associated with the optimized flip angle 87.9º after the first seven iterations.

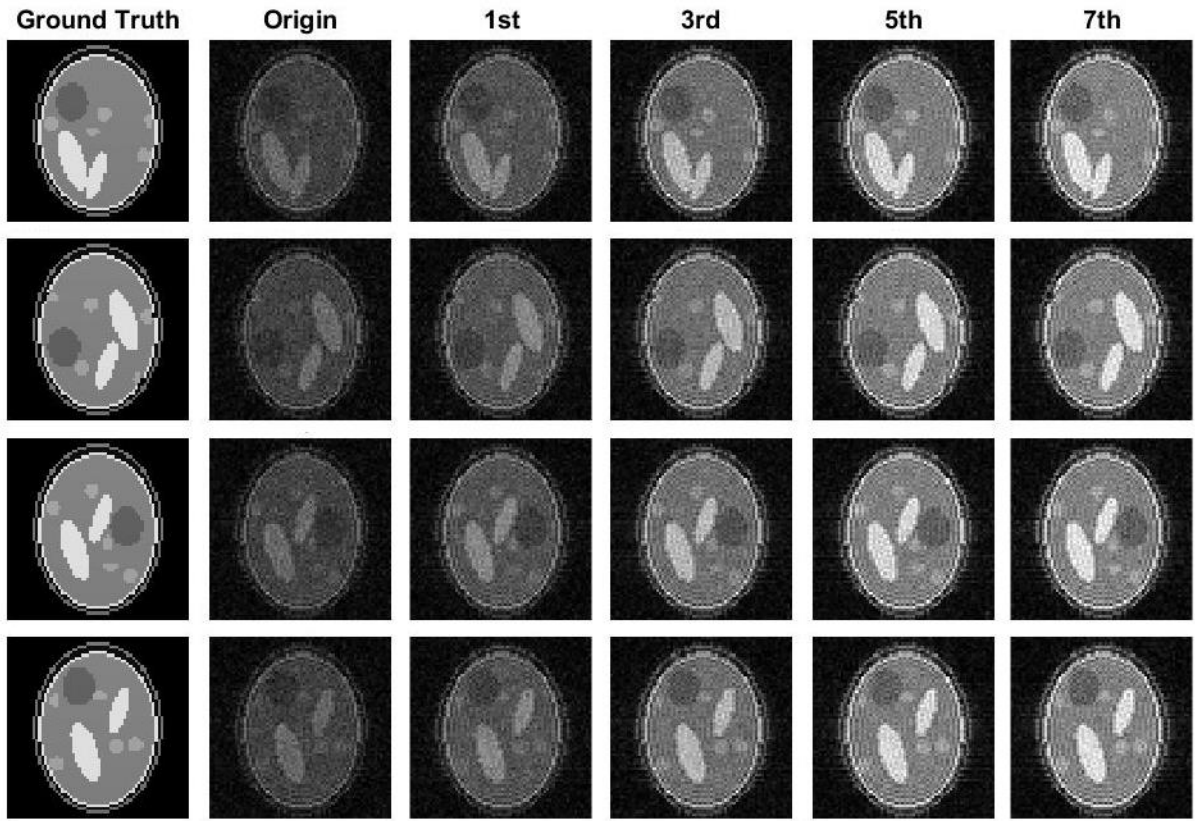

*Figure 6. Reconstructed MRI images through the Fourier inversion after the 1st, 3rd, 5th and 7th SPIN iteration*s.

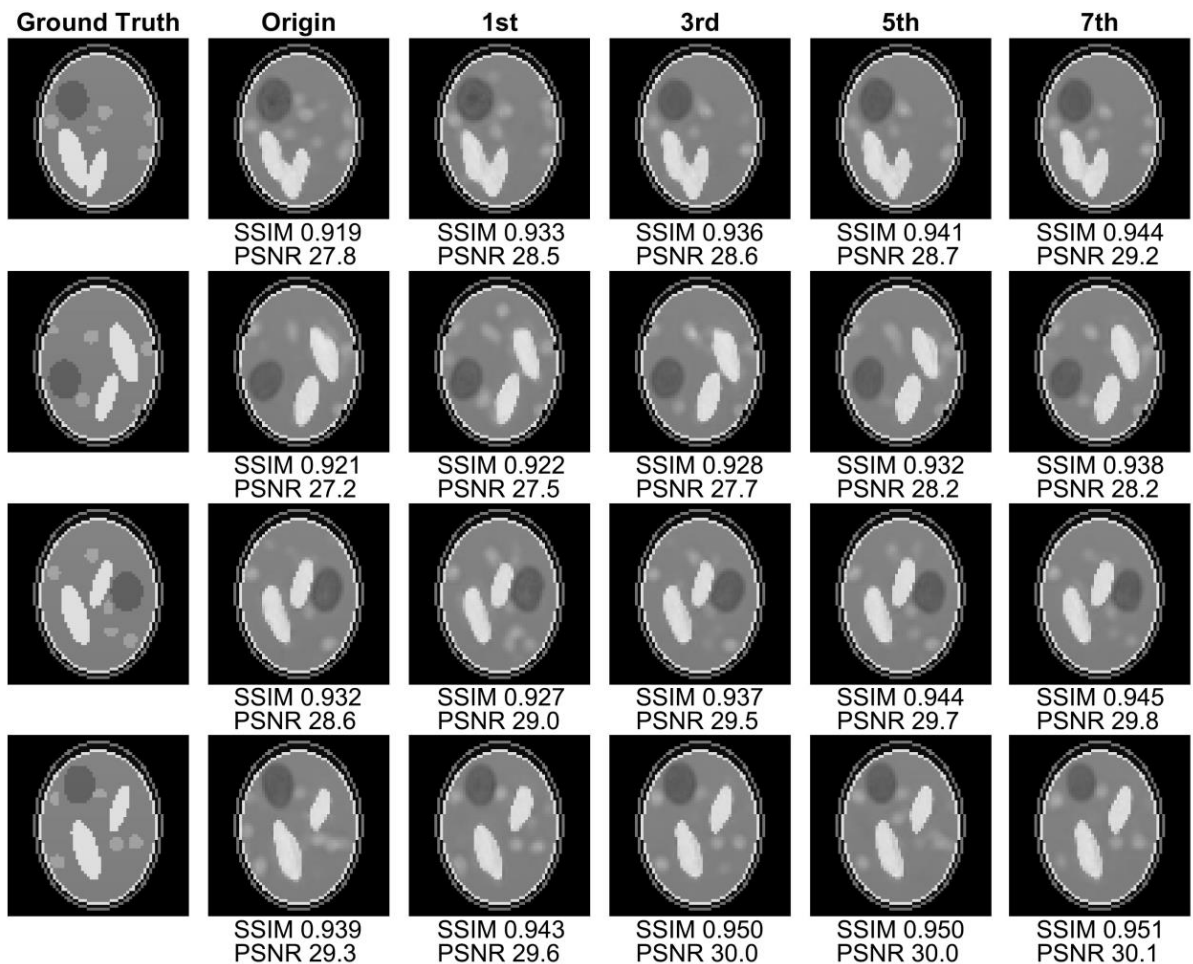

*Figure 7. Reconstructed MRI images with the reconstruction neural network after the 1st, 3rd, 5th and 7th SPIN iterations.*

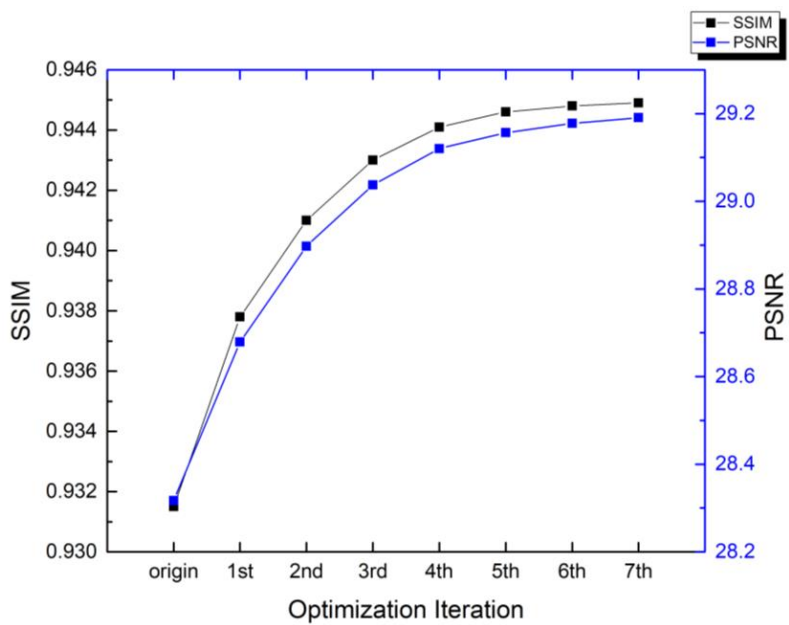

*Figure 8. Average SSIM and PSNR values of 1,000 reconstructed MRI images during the flip angle optimization process.*

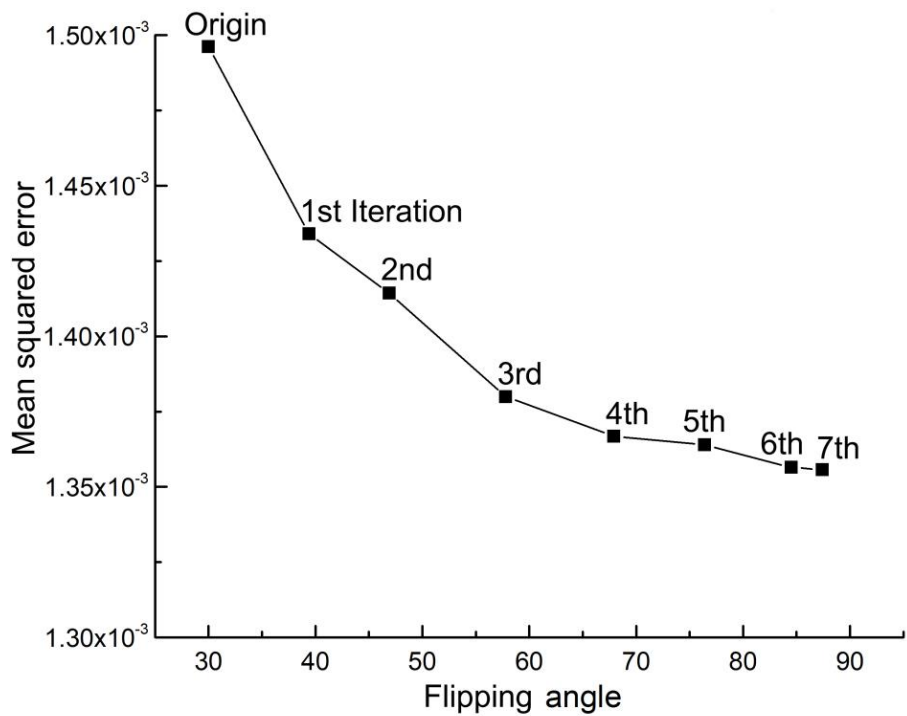

*Figure 9. Mean squared errors of the 1,000 reconstructed images after seven SPIN iterations. The horizontal and vertical axes are for the flip angle and the mean square error respectively. The plot shows the flip angle optimization after each iteration.*

## IV. Discussions and conclusion

For the first time, our SPIN approach attempts to directly design and optimize a pulse sequence in the machine learning framework. As the magnetization vectors are steered around for a maximized information content, the resultant data can be recorded, to a good degree, similar to randomized incoherent samples collected via compressed sensing. In the SPIN fashion, the *k*-space is naturally bypassed, avoiding any approximations for the traditional MRI formulation rooted in the *k*-space. Although our numerical demo in this paper is quite preliminary, the potential along this direction could be significant; for example, for task-based 3D MR fingerprinting imaging.

Although multiple simplifications were made in this pilot study for proof of concept, the idea of the SPIN strategy has been clearly demonstrated through an integrated optimization of the data acquisition and image reconstruction components. Further extensions should be made to handle realistic situations. For example, the discretized data acquisition process can be made more sophisticated by refining the time interval length and sampling step size. Also, the T2* effect can be included in the FID signal model if the pixel size is not sufficiently small [11]. The chemical shift and magnetic susceptibility could be taken into account as well [12]. More interestingly, MRI with an inhomogeneous background magnetic field can be naturally handled using our SPIN strategy. In the future studies, more advanced network architectures can be used for achieving better reconstruction results with higher resolution, such as U-NET [13], RED-CNN [14], and GAN [15].

SPIN is an intelligently streamlined MRI workflow from fundamental physics to first-hand information to optimized reconstruction regularized by big data, and finally can be also liked to radiomics features and final diagnosis. This is an example of the application of the systematic approach to MRI for precision medical imaging. In principle, the SPIN outcomes should be superior to that from a plain combination of a separately-performed pulse sequence design and an independently-optimized image reconstruction. As a next task, great efforts should be made to demonstrate real gains in preclinical and clinical tasks.

As a drawback of the SPIN approach, the overall MRI workflow demands a greatly-increased computational cost. We view this as opportunities to develop cost-effective algorithms, find highly-parallel computing engines such as GPU and TPU (and even quantum-computing algorithms), and analyze the involved non-convex optimization problem in terms of uniqueness, convergence, and stability. While SPIN is presented here for MRI, the idea can be extended to other imaging modalities and their combinations; such as ultrasound imaging and simultaneous emission-transmission tomography aided by polarized radiotracers [16].

In conclusion, we have proposed the SPIN strategy to synergize MRI data acquisition and image reconstruction in a unified machine learning framework, and reported simple but informative simulation results, illustrating key ingredients of SPIN to inspire further developments in the MRI field and beyond. Hopefully, SE, GR, EPI, fingerprint and other famous pulse sequences could be re-discovered by SPIN, and yet new pulse sequences be invented through machine learning as exemplified by SPIN. With SPIN-inspired data acquisition and network-based image reconstruction, it is possible that MRI will make a difference from the state of the art.

**REFERENCES**

[1] G. Wang, "A perspective on deep imaging," *IEEE Access,* vol. 4, pp. 8914-8924, 2016.
[2] C. M. Hyun, H. Pyung Kim, S. M. Lee, S. Lee, and J. K. Seo, "Deep learning for undersampled MRI reconstruction," *ArXiv e-prints arXiv: 1709*.02576, 2017
[3] S. U. Hassan Dar and T. Cukur, "A Transfer-Learning Approach for Accelerated MRI using Deep Neural Networks," *ArXiv e-prints arXiv: 1710.02615*, 2017
[4] D. Ma *et al.*, "Magnetic resonance fingerprinting," *Nature,* vol. 495, no. 7440, p. 187, 2013.
[5] H. M. Gach, C. Tanase, and F. Boada, "2D & 3D Shepp-Logan phantom standards for MRI," in *Systems Engineering, 2008. ICSENG'08. 19th International Conference on*, 2008, pp. 521-526: IEEE.
[6] B. Zhu, J. Z. Liu, S. F. Cauley, B. R. Rosen, and M. S. Rosen, "Image reconstruction by domain-transform manifold learning," *Nature,* vol. 555, no. 7697, p. 487, 2018.
[7] T. Tieleman and G. Hinton, "Lecture 6.5-rmsprop: Divide the gradient by a running average of its recent magnitude," *COURSERA: Neural networks for machine learning,* vol. 4, no. 2, pp. 26-31, 2012.
[8] M. Abadi *et al.*, "TensorFlow: A System for Large-Scale Machine Learning," in *OSDI*, 2016, vol. 16, pp. 265-283.
[9] X. Glorot and Y. Bengio, "Understanding the difficulty of training deep feedforward neural networks," in *Proceedings of the thirteenth international conference on artificial intelligence and statistics*, 2010, pp. 249-256.
[10] H. Shan *et al.*, "3D Convolutional Encoder-Decoder Network for Low-Dose CT via Transfer Learning from a 2D Trained Network," *IEEE Transations on Medical Imaging*, 2018
[11] G. B. Chavhan, P. S. Babyn, B. Thomas, M. M. Shroff, and E. M. Haacke, "Principles, techniques, and applications of T2*-based MR imaging and its special applications," *Radiographics,* vol. 29, no. 5, pp. 1433-1449, 2009.
[12] C. H. Ziener, "Susceptibility effects in nuclear magnetic resonance imaging," *ArXiv e-prints arXiv: 1703.04003*, 2017
[13] O. Ronneberger, P. Fischer, and T. Brox, "U-net: Convolutional networks for biomedical image segmentation," in *International Conference on Medical image computing and computer-assisted intervention*, 2015, pp. 234-241: Springer.
[14] H. Chen *et al.*, "Low-dose CT with a residual encoder-decoder convolutional neural network," *IEEE transactions on medical imaging,* vol. 36, no. 12, pp. 2524-2535, 2017.

[15] I. Goodfellow *et al.*, "Generative adversarial nets," in *Advances in neural information processing systems*, 2014, pp. 2672-2680.
[16] L. Gjesteby, W. Cong, and G. Wang, "Simultaneous Emission-transmission Tomography (SET)," in *The 14th International Meeting on Fully Three-Dimensional Image Reconstruction in Radiology and Nuclear Medicine*, Xi'an, 2017, vol. 14, pp. 363-371, Xi'an: Fully3D Community, 2017.